\documentclass[10pt,twocolumn]{article}
\usepackage{simpleConference}
\usepackage{times} 
\usepackage{subcaption}
\usepackage{svg}
\usepackage{scalefnt}
\usepackage{booktabs}
\usepackage{multirow}
\usepackage{csquotes}
\usepackage{xcolor}
\usepackage{colortbl}
\usepackage{tcolorbox}
\usepackage{listings}
\usepackage{array}
\usepackage{tikz}
\usepackage{algorithm}
\usepackage{graphicx}
\usepackage{amssymb}
\usepackage{xurl,hyperref}
\usepackage[noend]{algpseudocode}

\newcommand{\smallurl}[1]{{\scriptsize \url{#1}}}

\newtcolorbox{infobox}[2][]
{
	colframe = gray!25,
	colback  = gray!10,
	coltitle = gray!20!black,  
	title    = #2,
	#1,
}

\definecolor{codegreen}{rgb}{0,0.6,0}
\definecolor{codeblue}{rgb}{0,0,0.6}
\definecolor{codegray}{rgb}{0.5,0.5,0.5}
\definecolor{codepurple}{rgb}{0.58,0,0.82}
\definecolor{backcolour}{rgb}{0.95,0.95,0.92}
\definecolor{lightgray}{gray}{0.9}

\definecolor{lightgray}{RGB}{217,217,217}
\definecolor{lightlightgray}{RGB}{235,235,235}

\newcommand{\lightlightgrayrow}{\rowcolor{lightlightgray}}

\lstdefinestyle{mystyle}{
	backgroundcolor=\color{backcolour},   
	commentstyle=\color{codegreen},
	keywordstyle=\color{magenta},
	numberstyle=\tiny\color{codegray},
	stringstyle=\color{codepurple},
	basicstyle=\ttfamily\scriptsize,
	breakatwhitespace=false,         
	breaklines=true,                 
	captionpos=b,                    
	keepspaces=true,                 
	numbers=left,                    
	numbersep=4pt,                  
	showspaces=false,                
	showstringspaces=false,
	showtabs=false,                  
	tabsize=2,
	xleftmargin= 6pt,
}

\begin{document}

\title{An Approach to Build Consistent Software Architecture Diagrams Using Devops System Descriptors}

\author{
	Jalves Nicacio\\
	Université du Québec à Chicoutimi \\
	Chicoutimi - CA \\
	\texttt{\small jalves.mendonca-nicacio1@uqac.ca}
	\and
	Fabio Petrillo\\
	École de Technologie Supérieure - ÉTS \\
	Montreal - CA\\
	\texttt{\small fabio.petrillo@etsmtl.ca}
}

\maketitle
\thispagestyle{empty}

\begin{abstract}
	
System architecture diagrams play an essential role in understanding system architecture. They encourage more active discussion among participants and make it easier to recall system details.
However, system architecture diagrams often diverge from the software. As a result, they can interfere with the understanding and maintenance of the software.
We propose an approach to build system architecture diagrams using DevOps system descriptors to improve the consistency of architecture diagrams. 
To produce our approach, we survey problems with architecture diagrams in the software industry, developing guidelines for creating architecture diagrams.
Next, we produce a taxonomy for system descriptor concepts and a process to convert system descriptors into architecture diagrams.
We evaluate our approach through a case study. In this case study, we defined a Docker Compose descriptor for a newsfeed system and transformed it into a system architectural diagram using the proposed approach. 
Our results indicate that, currently, system descriptors generally lead to consistent diagrams only to a limited extent. However, the case study's observations indicate that the proposed approach is promising and demonstrates that system descriptors have the potential to create more consistent architectural diagrams.
Further evaluation in controlled and empirical experiments is necessary to test our hypothesis in more detail.

\end{abstract}



\section{Introduction}
Nowadays, we see a growing demand for software development that is increasingly complex and difficult to maintain. Moreover, modern software tends to be more performant and deals with quality issues such as scalability, security, maintainability, and internationalization.

For this reason, according to Fairbanks  \cite{fairbanks2010}, software systems are arguably the largest and most complex artifacts ever built. To mitigate software's scale and complexity issues, developers adopt processes and tools, such as software architectural modeling and, in this context, software architecture diagrams.

Software architecture diagrams are helpful for understanding and communicating about systems, as they encourage more active discussion among participants and make it easier to remember system details \cite{jolak2020}. In this sense, software architecture diagrams play a significant role in facilitating system architecture understanding.

However, software architecture diagrams (such as UML deployment diagrams) often diverge from the software. As a result, they can adversely affect the understanding and maintenance of the software. 
It is a real challenge to validate the consistency of diagrams for modeling software, and several research papers have presented different approaches to verifying the consistency of models. Some recurring approaches are: (1) validation of UML models using consistency rules \cite{Ha2003, Amor2011}, (2) algorithmic approaches \cite{litvak2003}, and (3) the use of graphs \cite{Mohammadi2014}.
These studies validate a UML model against another UML model. For example: Checking consistency between UML sequence and state diagrams \cite{litvak2003}, between UML structural diagrams and behavioral diagrams \cite{Ha2003}, between UML component and deployment diagrams \cite{Mohammadi2014}, and so on. Furthermore, even if the models are consistent (one with each other), there is no guarantee that these models will be consistent with the system in production. An alternative to ensure consistency of architecture diagrams with the system in production is to link them to key concepts related to continuous delivery (DevOps).

In the DevOps context, companies model and automate the high-level architecture of their systems through system descriptors. System descriptors are \textit{infrastructure as code} scripts \cite{morris2016} to automate application configuration management and deployment \cite{wurster2019}. For example, we can list AWS, Chef, CloudFormation, Docker Compose, Kubernetes, Puppet, and Terraform as system descriptors technologies. 

Thus, software architecture and DevOps share some concerns, such as security, scalability, and system monitoring \cite{bass2015}. Additionally, adopting systems descriptors proved to be a prominent approach to increasing the quality and productivity of information systems.

In light of the above, we propose an approach to build software architecture diagrams using DevOps system descriptors to improve the consistency of architectural diagrams. The main research question to be addressed is: \textit{Does creating software architecture diagrams using system descriptors improve the consistency of architecture diagrams?}

To answer our research question, we report a survey on problems with architecture diagrams in the software industry, developing a set of guidelines for creating architecture diagrams. These guidelines were beneficial in the context of this work, as they helped in the qualitative assessment of the diagram produced in our approach. Practitioners can also benefit from the guidelines by understanding what qualities to expect in an architectural diagram.

Applying our guidelines, we proposed our approach and developed a case study to evaluate it. In the case study, we defined a Docker Compose descriptor for a newsfeed system and transformed it into a system architectural diagram using our approach. Finally, we compare the original and the generated diagram to evaluate the obtained results.

The contributions of the paper are the following. (i) A meta-descriptor for transforming system descriptors into architecture diagrams of distributed systems; (ii) An approach to designing consistent architecture diagrams;  (iii) A set of guidelines to produce consistent architectural diagrams. Hence, the novelty of this work is the fact that it uses descriptors to generate architectural diagrams.

Our results indicate that, at the moment, system descriptors generally lead to consistent diagrams only to a limited extent. However, the case study's observations indicate that the proposed approach is promising and demonstrates that system descriptors have the potential to create consistent architectural diagrams.

We organized the remainder of this paper as follows: Section \ref{sec:Background-motivation} introduces the main concepts. Section \ref{sec:related-work} presents the related work. Then, section \ref{sec:survey} presents a survey on the architectural diagrams problems. Next, the section \ref{sec:approach} introduces our approach. We describe in Section \ref{sec:case-study} a case study on architectural diagrams generated from system descriptors. We discuss the obtained results in section \ref{sec:discussion}. Finally, Section \ref{sec:conclusions} concludes the article.


\section{Background}
\label{sec:Background-motivation}
This session provides some information about the theoretical framework linked to this work and background information about the tools and technologies used to implement the proposed approach.

\subsection{Architectural Diagrams}
A software architecture diagram is a graphical representation that shows how the elements of a system interact in a more extensive process. In particular, they help provide an overview and context of the system.

\begin{figure}[!t]
	\centering
	\includegraphics[width=.8\columnwidth]{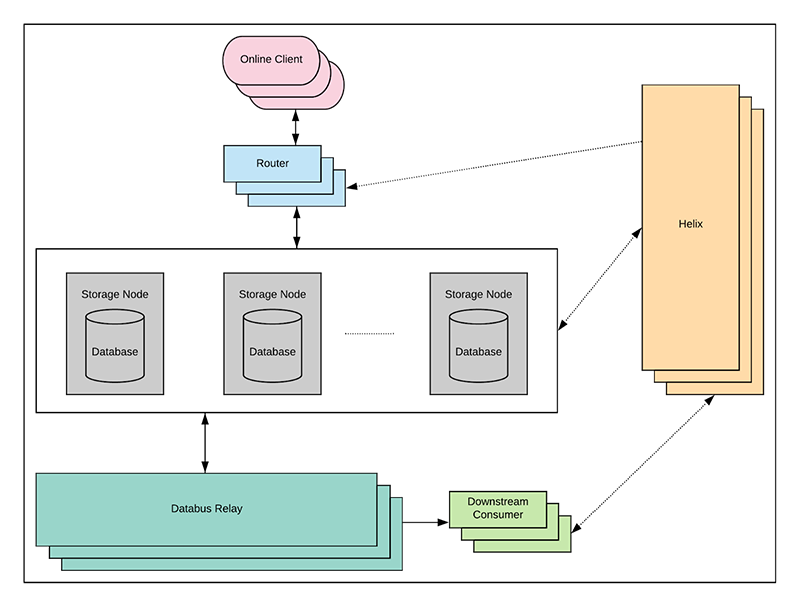}
	\caption{\textit{Ad hoc} architectural diagram example. Figure from  \cite{zhang2020}}
	\label{fig:uml-diagrams-examples}
\end{figure}

The UML has two principal diagrams related to software architecture: component diagram and deployment diagram. In addition to UML, other efforts have been to advance software architecture, such as SysML and ArchiMate. However, according to \cite{brown2021}, many development teams have already abandoned these notations in favor of simple "box and line diagrams" (\textit{ad hoc} diagrams). Figure \ref{fig:uml-diagrams-examples} shows an \textit{ad hoc} architecture diagram of a system.

According to \cite{jolak2020}, a diagram is better than a text description of the software to promote active discussion among developers about the details of the project. There are two striking features in the diagrams: (1) Architectural diagrams help to understand the system, and (2) Architectural diagrams improve communication and collaboration \cite{jolak2020}.

\paragraph{\textbf{Inconsistency in architectural diagrams}}
Litvak \textit{et al.} \cite{litvak2003} states that a consistency problem can arise because more than one diagram can describe the same aspects of the model.

To illustrate an example of inconsistency in architectural diagrams, consider the diagram in Figure \ref{fig:uml-diagrams-examples}. Analyzing this diagram, we could formulate several questions. For example, (1) how many services does this system have?; (2) where is this system deployed?; (3) how much does this diagram consistently communicate enough details to represent the same system using other languages or graphical notations, such as UML?

Thus, it is challenging to derive information about the system when certain visual elements are not present in the diagram. The idea of consistency in this paper refers to the understandability of the diagram. Thus, consistency is realized by how helpful the diagram is in understanding the system.

\subsection{System Descriptors}

System descriptors are scripts for automating, standardizing, and managing infrastructure in production environments.
System descriptors arise in the context of Infrastructure as Code (IaC), which uses configuration scripts to specify the definition and configuration of the software infrastructure required to run a system \cite{artac2017}.

In practice, system descriptors are artifacts that describe a system architecture. Tools such as Chef \footnote{Chef Infra: available at https://docs.chef.io/chef\_overview/}, Dockerfile\footnote{Dockerfile: available at https://docs.docker.com/engine/reference/builder}, and Puppet\footnote{Puppet: available at https://puppet.com/docs/puppet} create system descriptions. Container orchestrators such as Docker-Compose and Kubernetes \cite{dockercompose2019, kubernetes2020} also generate system descriptors.

According to \cite{huettermann2012}, system descriptors allow developers to treat system infrastructure in the same way as code development: choose the right tool and implement a solution that effectively meets the requirements. System descriptors are easily replicated, versioned, and computer-executable.


While these scripts are good at telling the computer accurate information about the architecture and infrastructure of the system, diagrams do a better job of communicating data to humans \cite{flaatten2020}.

\subsubsection{System Descriptors Key Concepts}
\label{sec:key-concepts}

Each system descriptor in the software industry uses its own terms and concepts to describe a system. The following sections presents the most important terms and concepts used by the most common system descriptors.

\paragraph{\textbf{Key concepts in Docker Compose}}

The key concepts in Docker Compose are services, volumes, network, image, and depends-on.

According to the Docker Compose documentation \cite{dockercompose2019}, \textbf{services} represent the containers to be created in the application. In addition, the service contains the necessary settings for the service to run, such as Docker image, environment variables, and depends-on directive. \textbf{Depends-on} expresses the dependencies between services at startup and shutdown, while the \textbf{image} directive specifies the image from which to start the container.



\paragraph{\textbf{Key concepts in Kubernetes}}

According \cite{kubernetes2020}, Kubernetes is a portable, extensible, open-source platform for managing containerized workloads and services. Kubernetes also facilitates both declarative configuration and automation. The main concepts in Kubernetes are pods, services, deployments.

\textbf{Pods} are the most minor deployable compute units developers can create and manage in Kubernetes. It is possible to create and manage multiple pods in a Kubernetes cluster through a workload controller such as \textbf{Deployment}. In this way, the Deployment workload controller automatically manages the number of pod replicas running on the system.


Kubernetes defines another concept for centrally managing access to pods: \textbf{services}. Thus, a service is an abstraction that defines a logical set of pods and a policy for accessing them \cite{kubernetes2020}.


\paragraph{\textbf{Key concepts in Terraform}}
According to \cite{terraform2022}, the key concepts in Terraform are provider, resource, resource module, infrastructure module, composition, and data source.

\textbf{Providers} are Terraform plugins used to interact with cloud providers, SaaS providers, and other APIs. Some examples of providers are: Amazon AWS, Google Cloud Platform, and Microsoft Azure. The Terraform settings must specify which providers are required for Terraform to install and use them.

Each provider adds a set of \textbf{resource} types or data sources that Terraform can manage. A resource belongs to a provider, accepts arguments, creates attributes, and has a lifecycle. A resource can be created, retrieved, updated, and deleted. Some examples of resources that belong to the aws provider are \textit{aws\_vpc}, \textit{aws\_db\_instance}, etc.




\section{Related Work}
\label{sec:related-work}

Concerning consistency verification in models, several research papers have presented different approaches to verifying the consistency of models. Some recurring approaches are: (1) validation of UML models using consistency rules \cite{Ha2003, Amor2011}, (2) algorithmic approaches \cite{litvak2003}, and (3) the use of graphs \cite{Mohammadi2014}. However, they proposed approaches to validate a UML model against another UML model. We proposed an approach that applies a valid system descriptor to generate architectural diagrams.

In addition, other related research focused on the transformation of system descriptors into graphical representations, such as \cite{piedade2020}, \cite{burco2020}, and \cite{sandobalin2019}.
One difference with our work is that we focus on improving architectural diagrams' consistency through system descriptors.



\section{A survey on the architectural diagrams problems}
\label{sec:survey}

In previous work\cite{nicacio2021}, when examining system descriptors and architectural diagrams in the literature, we found a lack of research on the quality of architectural diagrams in the software industry. We also found that engineers use general-purpose notations to create \textit{ad hoc} diagrams in tools such as Visio or draw.io\footnote{The tool is currently called Diagrams.net}.

Given this, we felt it was important to list guidelines that are easy to adopt and effective for improving the consistency and understandability of distributed systems architecture diagrams.

Therefore, in this session, we present a survey we conducted among software developers intending to list relevant guidelines for architectural diagram quality. The survey results are also significant for the context of this work since these guidelines lead to the quality of the architecture diagrams produced with our approach (Sec. \ref{sec:approach}).

We conducted a survey research in order to answer the following questions: 

\textbf{How inconsistent are distributed systems architectural diagrams with the system they represent?} This question aims to answer the impact of the architectural diagram inconsistency on the system's understandability and the ease of communication about the system components.

\textbf{Which aspects of architectural diagrams designed by the software engineering industry are positive and which are negative?}
We investigate what practitioners comment about the main problems with architectural diagrams of distributed systems.
The answer to this question helps researchers become aware of the difficulties and challenges that must be overcome when developing approaches for modeling distributed systems architecture. Also, this question helps practitioners identify the elements that act positively on the consistency and quality of architectural diagrams.

\textbf{What would be a set of best practices for designing distributed systems architectural diagrams?} The answer to this question aims to develop a set of practices that are easy to adopt and effectively improve the consistency of distributed systems architecture diagrams.

\subsection{Study Design}

We randomly selected five architectural diagrams from various technical blogs such as LinkedIn\footnote{\url{https://engineering.linkedin.com/blog/} }, Netflix\footnote{\url{https://netflixtechblog.com} }, Pinterest\footnote{\url{https://medium.com/pinterest-engineering/}}, Spotify\footnote{\url{https://labs.spotify.com/} }, and Twitter\footnote{\url{https://blog.twitter.com/engineering/en\_us/topics/infrastructure/}}. Figure \ref{fig:uml-diagrams-examples} shows one of five diagrams selected for study.

For each diagram in the questionnaire, we asked three free-text questions and five multiple-choice questions phrased on a five-point Likert scale \cite{jamieson2005} ranging from \textit{strongly disagree} to \textit{strongly agree }. 

We designed the questionnaire questions according to Unhelkar's model \cite{unhelkar2005}. We evaluated the quality of the diagrams by checking the syntax, semantic, and pragmatic (or comprehensibility) quality categories.

The scope of the survey only applies to people familiar with software development. Thus, we identified two profiles of participants in our study related to software development:  professional and student. Furthermore, to ensure that participants can provide reliable answers to the questions in the questionnaire, we have defined participant exclusion criteria if they are not related to the field of software engineering.

We invited a total of 132 people. Of these, 34 responded to the questionnaire, and 33 started the survey but did not complete it. The other invited individuals did not respond to the invitation to participate in the survey. 

Surveys usually have low response rates. According to Kitchenham, Budgen and Brereton \cite{kitchenham2015}, a score of 10\% is generally considered very good. Also, according to \cite{kitchenham2015}, the survey must have at least 30 responses to produce any form of statistics. We have 34 participants with a response rate of 25.7\%. So the number of participants made this survey possible.

The recruitment result was a sample composed mainly of mature, experienced respondents occupying a broad professional position in software engineering. The respondents were mainly from the United States and Canada, but there were also informants from Brazil. For this reason, we make the questionnaire available online in three different languages: English, French, and Portuguese.

\begin{figure}[t]
	\centering
	\includegraphics[width=0.6\columnwidth]{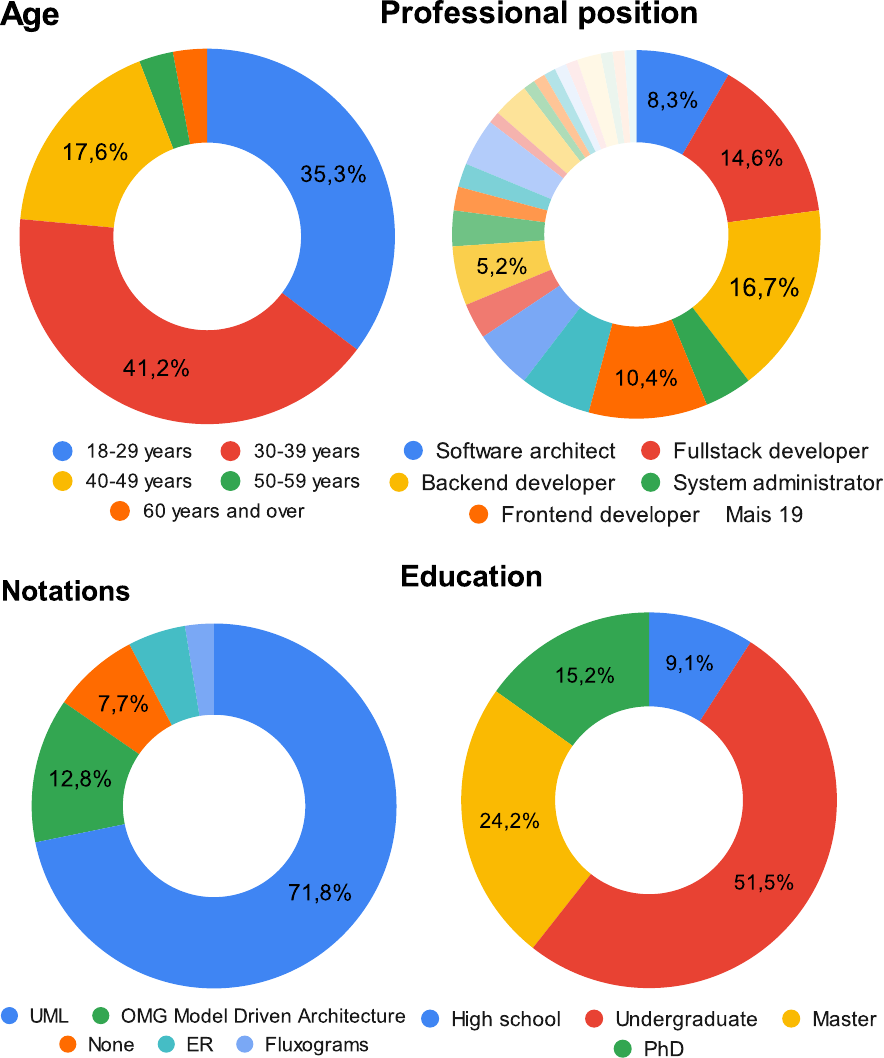}
	\caption{Demographic data}
	\label{fig:demographic-data}
\end{figure}

Figure \ref{fig:demographic-data} presents the main results of the demographic analysis of the respondents and shows that the sample is suitable for answering the survey questions. The sample consists mainly of people between the ages of 30 and 39 ($41.2\%$). In addition, $51.5\%$ of respondents are undergraduates, and $39.4\%$ of respondents are second and third-cycle graduates ($24.2\%$ Master's and $15.2\%$ Ph.D.).

\subsection{Survey Results}
\label{sec:survey-results}

\paragraph{Diagrams consistency}

\begin{figure}[h]
	\centering
	\includegraphics[width=0.8\columnwidth]{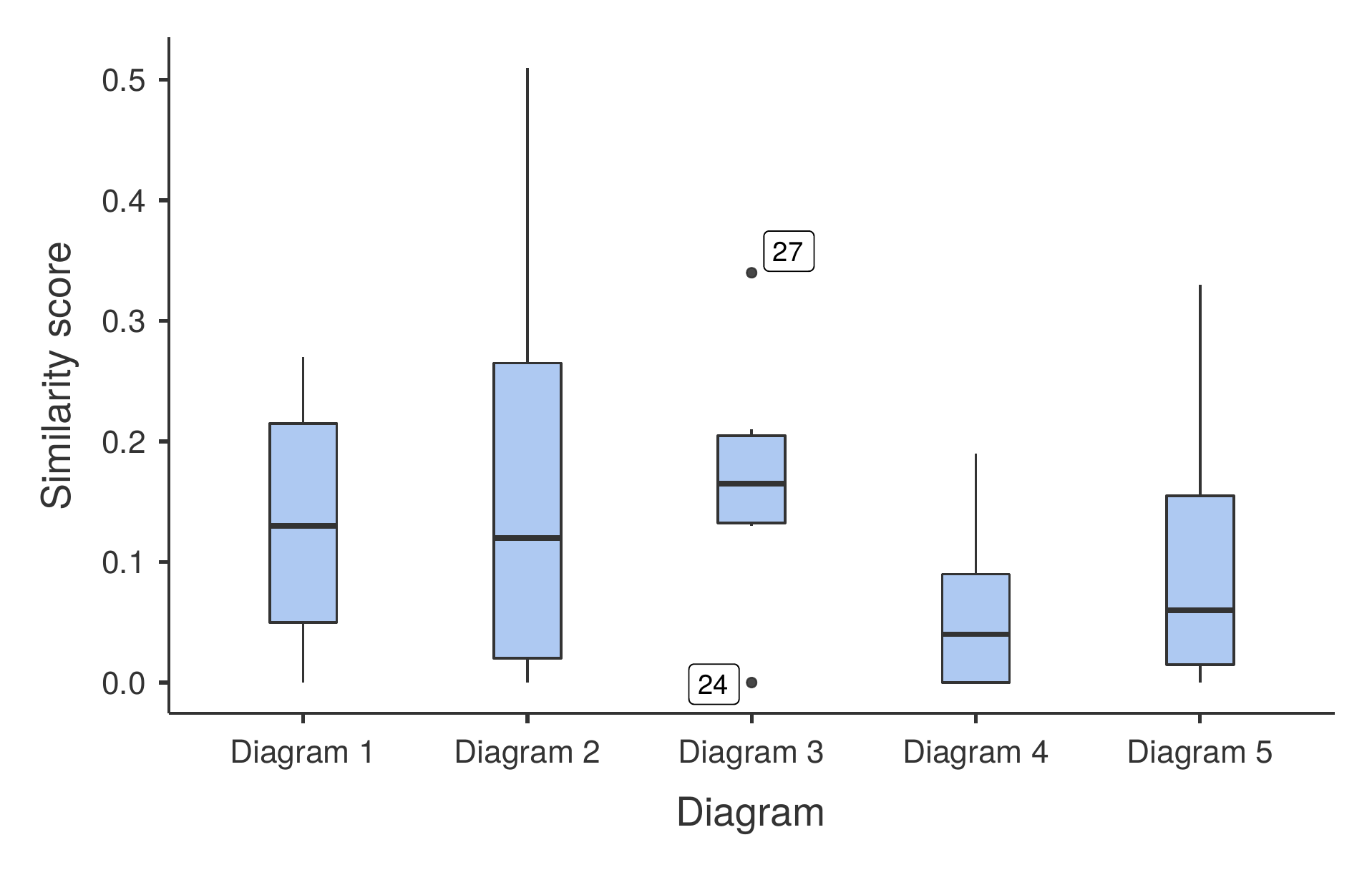}
	\caption{{\tiny The similarity between the answers to the first question of the questionnaire and the original texts of the architectural diagrams}}
	\label{fig:q1}
\end{figure}

We asked participants to describe the system represented by the diagram. We then analyzed the responses to measure the similarity score between the participants' responses and the original text. By doing so, we want to measure how far the participant's answer is from the author's intention. We used the cosine similarity metric to calculate the similarity score between the participants' texts and the original text, using a scale from 0 to 1. Figure \ref{fig:q1} shows the distribution of the similarity value in a boxplot.

Looking at the figure \ref{fig:q1} and evaluating the degree of dispersion of the answers for each diagram, we notice that the answers for diagram 2 show a more significant variability of similarity scores. On the other hand, diagram 3 has the highest median and the lowest dispersion of similarity scores among all diagrams. Nevertheless, the responses for diagram 3 have a low similarity score.

In conclusion, the participants' responses bear little resemblance to the original texts. \textbf{This fact suggests that the diagrams do not have enough elements to ensure consistency with participants' responses.}

\paragraph{Diagram strengths and weaknesses}

We asked the participants to list the strengths and weaknesses of the architectural diagrams presented in the survey according to their opinion.

\begin{table}[h]
	\caption{List of keywords for the category "Using shapes and arrows" and some extracts from answers}
	\label{tab:keyword-category}
	\begin{tabular}{@{}ll@{}}
		\toprule
		\multicolumn{1}{c}{Keyword} & \multicolumn{1}{c}{Answer extracts}             \\ \midrule
		Shapes                      & \enquote{Geometric shapes}   \\
		Arrows                      & \enquote{Arrows to indicate the relationship }                                       \\
		Different types of arrows   & \enquote{The different types of arrows}                                        \\
		Variety of shapes           &                                           \\ \bottomrule
	\end{tabular}
\end{table}

We extracted keywords from the participants' responses to analyze the collected data. We then categorized the keywords and calculated how frequently each category appears in the responses. For example, the table \ref{tab:keyword-category} shows the keywords grouped into the category "use of shapes and arrows," as well as some excerpts from the responses that illustrate how we extracted the keywords.


The emerging conclusion from this data is that \textbf{designers generally make the most significant mistakes on the elements identified as the most important in a diagram.}

\paragraph{Syntax, semantics, and comprehension}

In the last question, we asked each participant to classify five statements according to their opinion. The statements through a scale from "strongly disagree" to "strongly agree." Thus, we used the Likert scale to interpret the data for this question.

In summary, participants tend to agree that the diagrams generally use the correct notation and that the elements are correctly labeled with the appropriate labels. However, 41\% of participants disagreed or strongly disagreed that "all expected elements are present in the diagram," and 38\% neither agreed nor disagreed with this statement. From the analysis of the responses, we can infer that \textbf{participants lack more details to adequately interpret and understand the diagram}.

From the observations of the data collected in the survey, we suggest guidelines to produce consistent architectural diagrams are: (1) The elements of the diagram must be clearly identified; (2) The notation used in the diagram must be clearly defined; (3) The elements of the diagram must be presented in more detail; (4) The diagram must define a starting and finishing point for reading and interpreting the diagram and (5) The diagram must adopt symbols for specific technologies if the system uses those technologies.

\section{Approach}
\label{sec:approach}

We propose the use of system descriptors to improve the consistency of architectural diagrams, to build and validate them. Our hypothesis is that \textbf{if an architectural diagram is generated from a valid system descriptor, then the diagram is consistent.}

We assume that a descriptor file is inherently consistent for two reasons: (1) system descriptor files are written in a formal language, such as YAML (used by Docker-compose and Kubernetes pods) or Ruby (used by Chef); (2) system descriptor files are automatically processed by a finite state machine. 
In this way, if the diagram generated from the system descriptors represents each descriptor element, this diagram is also consistent.

There are several technologies for system descriptors that support different functions and use their modeling languages. Therefore, it is not easy to compare technologies based on their capabilities \cite{wurster2019}.

Since we intend a technology-independent approach to system description transformation, we present a taxonomy of system descriptor concepts in Section \ref{sec:taxonomy}. We based the terminology used in this taxonomy on the Essential Deployment Meta-Model (EDMM) proposed by Wursters \textit{et al} \cite{wurster2019}. Finally, in Section \ref{sec:meta-descriptor}, we introduced a meta-descriptor and a transformation function for system descriptors.

\subsection{A Taxonomy for System Descriptor Concepts}
\label{sec:taxonomy}
Since we intend to build architectural diagrams from a system descriptor, we must first define a set of terms representing the distributed system's elements. For example, the UML deployment diagram uses the terms nodes for each type of hardware and artifacts for software components \cite{rumbaugh2004}.

However, each system descriptor uses its own terms and concepts to describe a system. Moreover, different system descriptors sometimes use the same terminology but different meanings. For example, Docker Compose uses the term "services" to describe a compute resource within an application that can be scaled or replaced independently of other components \cite{dockercompose2019}. Kubernetes, on the other hand, uses the terminology "service" to describe an abstraction that defines a logical set of pods and a policy for accessing them \cite{kubernetesservice2022}.

According to the meta-model proposed by Wurster \textit{et al}. \cite{wurster2019}, the essential elements for the deployment of system architecture are components and relations. In addition, component and relation types specify the semantics of these elements.

Therefore,  we propose a taxonomy based on observing the terminology used in Docker Compose, Kubernetes, and Terraform system descriptors. Section \ref{sec:key-concepts} presented the key concepts for these system descriptors. 

\newcommand{\centered}[1]{\parbox[m]{2.4cm}{#1 \vspace*{1pt}}}
\begin{table}[h]
	\scalefont{0.7}
	\caption{Taxonomy for modeling system architectures based on system descriptor terminology}
	\label{tab:taxonomy}
	\begin{tabular}{m{2.2cm} p{1.2cm}p{1.8cm}p{1.8cm} }
		
		\toprule
		
		\multicolumn{1}{c}{Categories} &
		\multicolumn{1}{c}{Compose} &
		\multicolumn{1}{c}{Kubernetes} & 
		\multicolumn{1}{c}{Terraform} \\ 
		
		\midrule
		
		\centered{Components} & Services & Pods, Deployment & Resource, modules \\
		\lightlightgrayrow \centered{Components propertyes} & Volumes, ports, expose & Volumes, targetPort, port  & Data source, Input Variables    \\
		\centered{Component type} & - & labels & Resource type  \\
		\lightlightgrayrow \centered{Dependency relation} & Depends\_on & - & Depends\_on    \\
		\centered{ Infrastructure nodes } & Deploy & Service, Deployment, Node & Resource, Composition, Infrastructure module, Resource module  \\
		\lightlightgrayrow \centered{Artifacts} & Image, build & Image & Provider, resource type  \\
		\bottomrule
	\end{tabular}
\end{table}

We classify terms according to their purpose. For example, each descriptor uses a different term to represent its minimal deployable unit: Service (in Docker Compose), Pod (in Kubernetes), and Resource (in Terraform). In this way, we have classified the key terms into six different categories, as shown in the taxonomy in Table \ref{tab:taxonomy}.

\paragraph{\textbf{Components}} Components are deployable nodes that represent elements instantiated for system execution. They are associated with system features and business rules. Some examples of components are system modules or APIs, microservices, and databases.

\paragraph{\textbf{Component properties}} They are a set of information used to configure a component. Properties must describe either the current or desired state of a component. For example, to access files stored in locations outside the container, Docker Compose and Kubernetes use the term \textit{volume} to associate the data created and used by Docker containers. Terraform, on the other hand, uses the term \textit{data source}. Other examples of properties are \textit{Ports}, \textit{Networks}, \textit{Input Variables}, and \textit{Environment Variables}.

\paragraph{\textbf{Component types}} Wurster \textit{et al} defines component types as an entity that specifies the semantics of a component \cite{wurster2019}. In Kubernetes, for example, one can use key/value pairs called \textit{labels} to identify a component type. In Terraform, identifying the component type (so-called resource type in the Terraform context) is mandatory when creating a component. 
Further, Docker Compose does not provide component type information. One possibility would be for the developer to use the terms services and container\_name simultaneously to indicate the name and type of the component, respectively.

\paragraph{\textbf{Artifact}} According to \cite{wurster2019}, an artifact implements a component. Refer to the resource used to deploy system components. For example, a Docker Compose service is deployed on an application stack contained in a Docker image. Docker accesses this image via an image repository or can build it from a Dockerfile artifact.

\paragraph{\textbf{Dependency relation}} The dependency relation specifies a dependency when a system component or module is physically, logically, or functionally dependent on another component. For example, the dependency relation is valuable when you need to run processes in a specific order.

Docker Compose and Terraform use the term \textit{depends\_on} to specify relationships between components. 
Kubernetes, on the other hand, does not inherently support the use of a term like \textit{depends\_on} in its descriptor. Hence, it is only possible to identify relationships between components in Kubernetes at the level of environment variables or within pods.

\paragraph{\textbf{Infrastructure nodes}} Infrastructure nodes are nodes related to the infrastructure functions of the system. They support architectural decisions and software constraints. Infrastructure nodes include Load Balancers, Messaging Middleware, Firewalls, DNS Services, Security Groups, etc.

\subsection{From System Descriptors to Architecture Diagrams}
\label{sec:meta-descriptor}

We want to implement the transformation between two models: from system descriptors to system architecture diagram. According to Stevens \cite{stevens2008}, model transformations aim to make software development and maintenance more efficient. The model transformation process is accomplished by automating routine aspects of the process. Once the process is automated, a model transformation can capture the notion of consistency.

\begin{figure}[h]
	\centering
	\includegraphics[width=0.99\columnwidth]{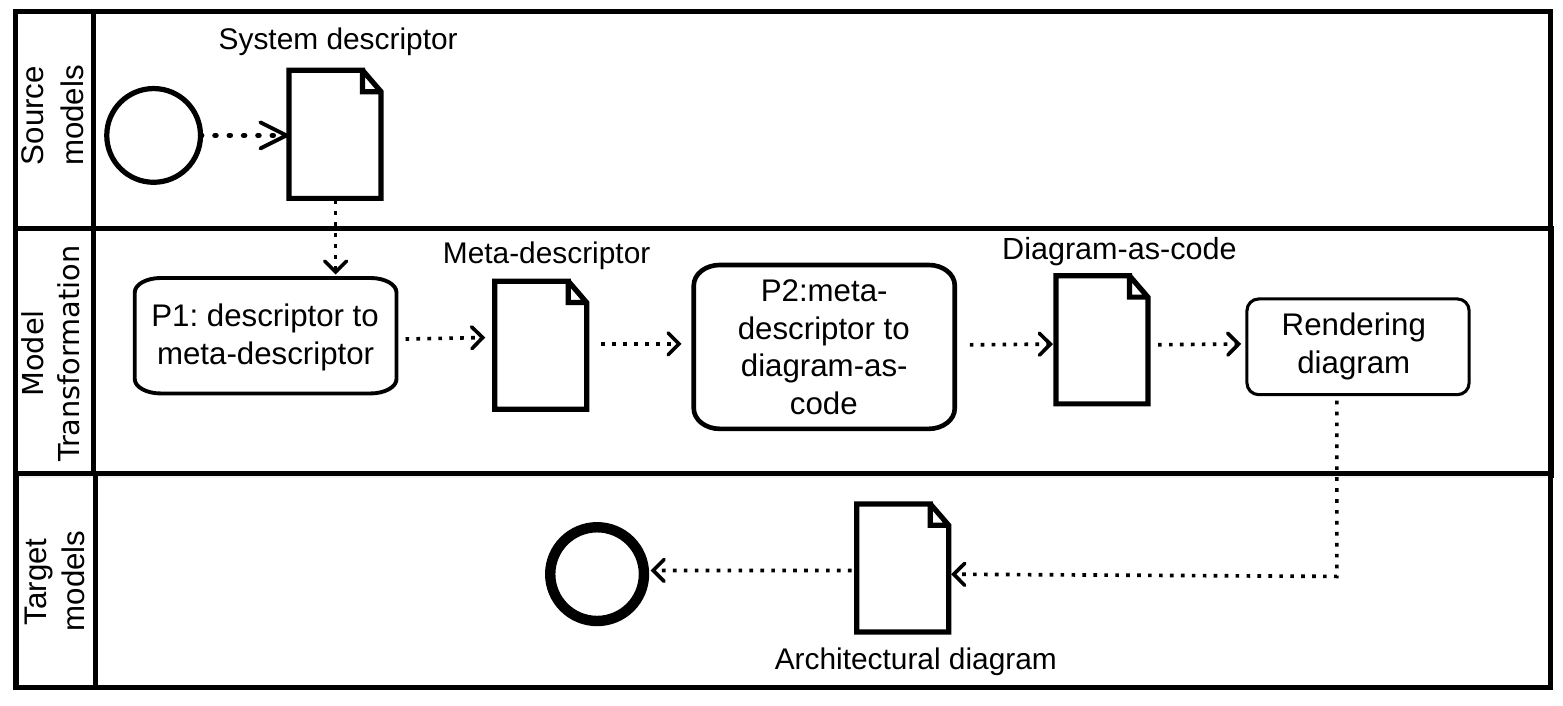}
	\caption{Transformation process}
	\label{fig:model-transformation}
\end{figure}

Figure \ref{fig:model-transformation} shows the transformation model process proposed in this paper. System descriptor files are source models at the beginning of the transformation process. System descriptors can be of different types, e.g., Docker Compose, Kubernetes,  and Terraform scripts.

The first step is to transform a given descriptor into a meta-descriptor (P1). Next, it transforms the meta-descriptor into a "diagram as code" model (P2). Finally, the rendering process  maps the diagram-as-code model to the architecture diagram model.

\paragraph{\textbf{Meta-descriptor into a diagram-as-code model}}
We used the taxonomy categories shown in Table \ref{tab:taxonomy} to create the meta-descriptor. Listings  \ref{lst:kubernetes-pod}, \ref{lst:docker-compose}, and \ref{lst:meta-descriptor} illustrate the P1 process. In this example, the descriptors of Kubernetes (Listing \ref{lst:kubernetes-pod}) and Docker Compose (Listing \ref{lst:docker-compose}) represent the same system: a simple API whose source code is stored in the \textit{emondek/simple-api} image\footnote{Node.js Simple REST API available at https://hub.docker.com/r/emondek/simple-api }. Therefore, in this example, the P1 process must always produce the script from Listing \ref{lst:meta-descriptor} regardless of the input descriptor.

\begin{lstlisting}[caption={Kubernetes pod script}, label={lst:kubernetes-pod}]
	apiVersion: v1
	kind: Pod
	metadata:
	name: simple-api
	labels:
	app: simple-api
	role: api
	spec:
	containers:
	- name: simple-api
	image: emondek/simple-api:1.0.0		
	ports:
	- containerPort: 3000
\end{lstlisting}

\begin{lstlisting}[caption={Docker Compose script}, label={lst:docker-compose}]
	services:
	simple-api:
	container_name: api
	image: emondek/simple-api:1.0.0
	ports:
	- "3000:3000"
\end{lstlisting}

\begin{lstlisting}[caption={Meta descriptor script}, label={lst:meta-descriptor}]
	components:
	simple-api:
	name: simple-api
	type: api
	properties:
	targetPort: 3000
	artifacts:
	image: emondek/simple-api:1.0.0
\end{lstlisting}

A meta-descriptor consists of a list of components and a list of relationships between components. A component can store information about its type, properties, and artifacts. A Docker image is an example of an artifact that contains the complete stack, starting from the operating system and ending with the application-specific component.

We use the YAML-based language for the meta descriptor, as this is a recognized standard among system description and deployment tools.

\paragraph{\textbf{Converting the meta-descriptor into a diagram-as-code model}} The P2 transformation process is responsible for converting the meta-descriptor into a diagram-as-code model. Diagram-as-code is an approach to generating diagrams through programming. This approach has sparked recent interest among software engineering practitioners. A non-exhaustive list of some web articles on the subject is given in \cite{flaatten2020, meyer2019, mingrammer2020, brown2020a}. 

The transformation algorithm for the P2 process depends on the DaC model chosen. In general, DaC models describe the components and their relationships. Additional information about these models can be: the display orientation of the components (right to left, top to bottom, etc.) and the export format of the diagram. 

Below, in Listing \ref{lst:meta-descriptor-to-dac}, we present a pseudo-algorithm that describes the main steps for the transformation between the meta-descriptor and the diagram-as-code.

\begin{lstlisting}[caption={Meta-descriptor to Diagram-as-Code algorithm}, label={lst:meta-descriptor-to-dac}]
	For each component in meta-descriptor components:
	Generate node representation code for a specific DaC:
	Add properties to DaC node
	Add component type to DaC node
	Add the artifacts to the DaC node
	For each relation in meta-Descriptor relations:
	Generate the arc representation code for a specific DaC:
\end{lstlisting}

\paragraph{\textbf{Rendering process}}
It is the process responsible for rendering the diagram from the diagram-as-code model. Nowadays, there are a number of text-based tools that generate diagrams, such as PlantUML\footnote{PlantUML is available at http://plantuml.com/}, Mermaid\footnote{Mermaid is available at https://mermaid-js.github.io/mermaid/}, Diagrams \footnote{Diagrams is available at https://diagrams.mingrammer.com/}, and Structurizr.

In the context of this work, we use Structurizr to generate the architectural diagrams. Structurizr is a tool for creating system architecture diagrams \cite{brown2022}. This software has its own domain language (DSL) for representing elements of the system architecture, which we can use to automatically create architecture diagrams.

\paragraph{\textbf{Meta-descriptor algorithm}} The algorithm for generating meta-descriptors may vary depending on the system descriptor used as input. For example, in Algorithm \ref{alg:meta-descriptor-algorithm}, we present the algorithm for converting a Docker Compose descriptor into a meta-descriptor.

Since Docker Compose consists of a single description file, a loop that goes through all the services is enough to allocate them. Therefore, the algorithm has two parts. The first part collects the information from each service and maps that information as properties of the meta-descriptor component. The second part checks all the relations for each service.

\begin{algorithm}
	\scriptsize
	\caption{Docker Compose to meta-descriptor algorithm}\label{alg:meta-descriptor-algorithm}
	\begin{algorithmic}
		\For{each service in Docker Compose services}
		\State Create a new $component$ with the following properties (if available): \Comment{part \#1}
		\State $name \gets service.name$
		\State $type \gets service.container\_name$
		\State $properties \gets [service.volumes, service.ports, ...]$
		\State $artifacts \gets get\_artifacts\_from\_service()$ \\
		
		\For{each $depends\_on$ in $service.depends\_on$} \Comment{part \#2}
		\State Create a new $relation$ with:
		\State  $out \gets service.name$
		\State  $in \gets depends\_on$ 
		\State Adds the new $relation$ to the list of relations
		\EndFor
		\EndFor \\
		
		\Function{get\_artifacts\_from\_service}{}
		\If{exists service.image}
		\State $artifacts.image \gets service.image$
		\ElsIf{exists service.build}
		\State $artifacts.image \gets service.build$
		\EndIf
		\If{exists service.deploy.replicas}
		\State $artifacts.replicas \gets service.deploy.replicas$
		\EndIf
		
		\State \Return $artifacts$
		\EndFunction
		
	\end{algorithmic}
\end{algorithm}

\paragraph{\textbf{Transformation function for system descriptors}}

It is possible to describe the models transformation as bidirectional or unidirectional functions \cite{stevens2008}. 

Let us assume that $\sigma$ is a script instance of a system descriptor, and $S$ is the set of all script instances. We also assume that $\delta$ is an architectural diagram instance and $D$ is the set of architectural diagrams. Hence, we define the transformation function as $f: S \rightarrow D$, so that:

\begin{equation}  
	f(S) = \{\sigma \in S \mid (\exists \delta \in D), f(\sigma) = \delta\} \subset D
\end{equation}
is the image of $f$. 

Thus, the transformation function must always find a diagram instance that matches each system descriptor. Moreover, we know from \cite{stevens2008} that the consistency relation is that $\sigma$ and $\delta$ are consistent if and only if $ \delta = f(\sigma)$.

\section{Case Study}
\label{sec:case-study}

In this session, we present a case study to evaluate our proposal. First, we describe a newsfeed system like Facebook or Instagram and show a high-level diagram depicting the system's architecture. We then implement system descriptors using Docker Compose. The newsfeed application described in \cite{xu2020} inspired the authors.

\subsection{A Newsfeed System Design}

A news feed is a list of newly published content on a website. End users can receive push updates for new content on a website by subscribing to the website's news feed. Xu \cite{xu2020} describes a newsfeed system in detail. 

Let us assume a newsfeed system like Facebook, Instagram, or Twitter. This newsfeed system contains two streams: feed publishing and newsfeed retrieval. The feed publishing stream allows users to publish posts, and the retrieval stream summarizes friends' posts in reverse chronological order. The newsfeed system stores the corresponding post data in the database or in the cache.

The user creates a post via the publishing service, where a news feed publishing API takes the user's authentication details and the post data (text, photos, or videos). Then a \textit{fanout service} pushes the new content to friends' news feeds. We call \textit{fanout service} the process of forwarding a post to all followers. 

\begin{figure}[htp]
	\includegraphics[width=0.97\columnwidth]{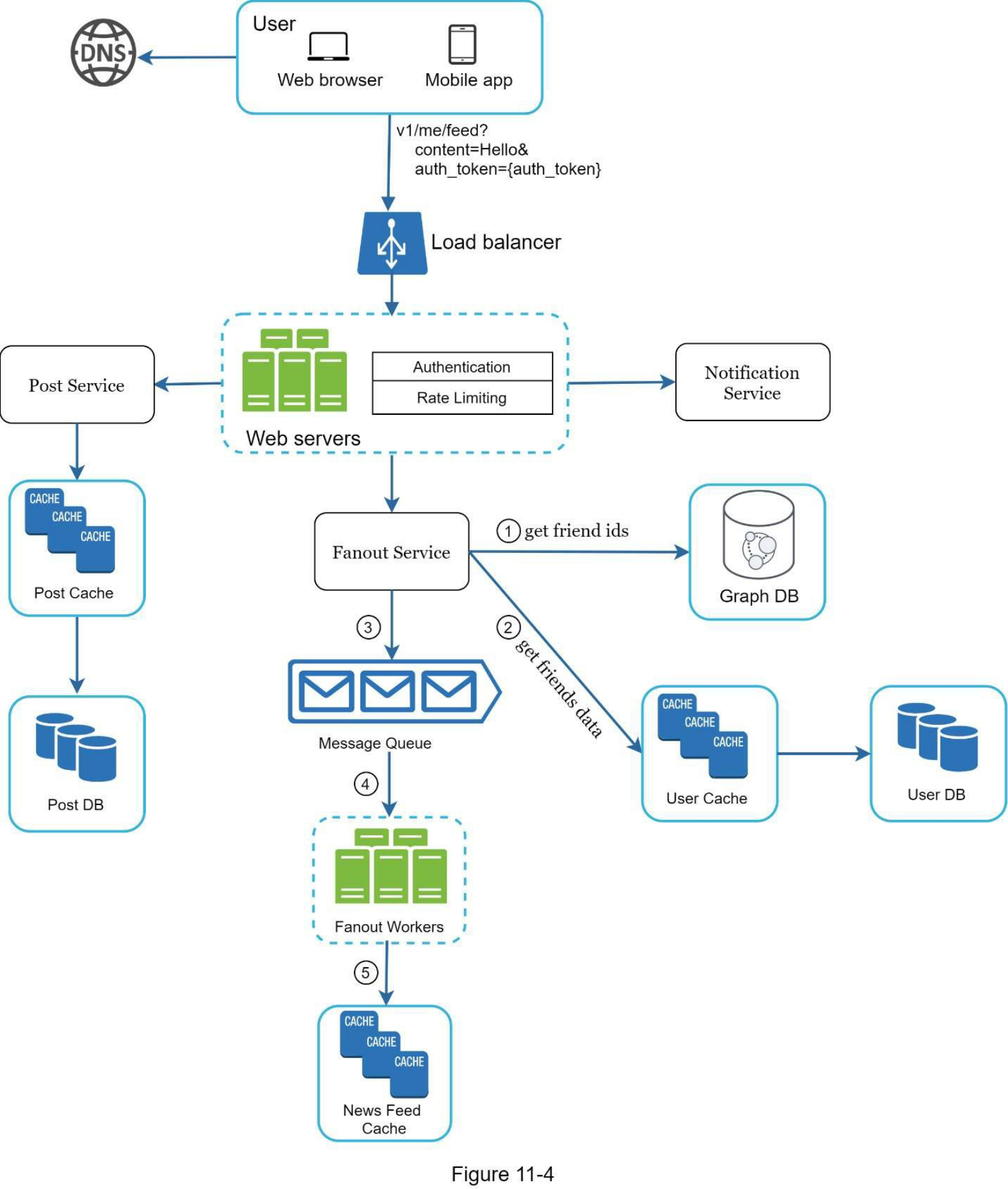}
	\caption{High-level architecture of a newsfeed system. Diagram from \cite{xu2020}}
	\label{fig:newsfeed-system-adhoc-xu-diagram}
\end{figure}

A modeling suggestion for this system is to design it as in Figure    \ref{fig:newsfeed-system-adhoc-xu-diagram}, 
which shows the high-level system architecture. The system components are as follows:
\begin{itemize}
	\item Load Balancer: distributes the load of requests among the web servers.
	\item Web server: maintains connections with users, forwards traffic to various services, and performs user authentication.
	\item Post Service: takes care of persistence of user posts in the database and cache.
	\item Fanout service: forwards new content to the feed of friends and followers.
\end{itemize}

The Fanout service uses a communication structure based on the publisher/subscriber pattern. In this context, it is widespread for the developer to use an implemented solution to satisfy the need for message exchange between system components. Some market-ready platforms that implement this concept of producer and consumer are Apache Kafka, ActiveMQ, and Redis.

\subsection{A Newsfeed System Descriptor}
Using the description and the architectural diagram, we created a Docker Compose descriptor. Listing \ref{lst:docker-compose-newsfeed-system-descriptor} depicts the system components according to the Docker Compose description language. Due to space limitations, we present only part of the descriptor in this section. Readers interested in more details can find the full version in our replication package\footnote{Repository available at https://github.com/jalvesnicacio/newsfeed-system-case-study}.

Note that the \textit{nginx service} (see line 2) has the role of load balancing according to the system description, although the Docker Compose descriptor does not contain this information. This issue affects the creation of the meta-descriptor, as it is not possible to assign missing information.

Table \ref{tab:docker-compose-services-for-newsfeed-system} also summarizes all Docker services created for the newsfeed system and some important aspects of these services. So we see that Docker Composer has eight services specified. Moreover, some of these services are linked via the \textit{depends\_on} property, creating a dependency relationship between them.

\begin{lstlisting}[caption={Docker Compose newsfeed system descriptor}, label={lst:docker-compose-newsfeed-system-descriptor}]
	services:
	nginx:
	image: nginx:latest
	restart: always
	volumes:
	- ./nginx.conf:/etc/nginx/nginx.conf:ro
	depends_on:
	- webserver
\end{lstlisting}

\begin{table}[]
	\caption{Docker Compose services for newsfeed system}
	\label{tab:docker-compose-services-for-newsfeed-system}
	\resizebox{\columnwidth}{!}{%
		\begin{tabular}{@{}llll@{}}
			\toprule
			Services        & Image                     & Depends\_on & Deploy      \\ \midrule
			nginx           & nginx:latest              & webserver   &             \\
			webserver       & emondek/simple-api:latest &             & replicas: 3 \\
			db              & postgres:14.1-alpine      &             &             \\
			cache           & redis:6.2-alpine          &             &             \\
			postservice     & emondek/simple-api:latest & db, cache   &             \\
			fanoutservice   & emondek/simple-api:latest & db, cache, redis   &        \\
			redis           & redis:6.2-alpine          &             &              \\
			worker          & app-image:latest          & redis       & replicas: 3  \\ \bottomrule
		\end{tabular}%
	}
\end{table}

\subsection{Transformation P1}
If we use the system descriptor in Listing \ref{lst:docker-compose-newsfeed-system-descriptor} as input to the algorithm in Algorithm \ref{alg:meta-descriptor-algorithm} we get the meta-descriptor in Listing \ref{lst:newsfeed-system-meta-descriptor}. This task corresponds to process P1 in our approach.

We show only a fragment of the code generated for the meta-descriptor in Listing \ref{lst:newsfeed-system-meta-descriptor}. Comparing the \textit{nginx} service from Listing \ref{lst:docker-compose-newsfeed-system-descriptor} with the \textit{nginx} component of the meta-descriptor, we see that the algorithm capture the essential elements for generating the architecture diagram.

\begin{lstlisting}[caption={Newsfeed System meta-descriptor}, label={lst:newsfeed-system-meta-descriptor}]
	components:
	nginx
	name: nginx
	properties:
	volumes:
	- ./nginx.conf:/etc/nginx/nginx.conf:ro
	artifacts:
	name: nginx:latest
	
	relations: 
	- out: nginx
	in: webserver
\end{lstlisting}

\subsection{Transformation P2 and Rendering}
We have chosen to use an automated tool for rendering the diagram. For this reason, the P2 process of our approach generates a diagram-as-code artifact that corresponds to the DSL language of the Structurizr tool. Listing \ref{lst:newsfeed-dac-model} presents a fragment of the Diagram-as-code model.

\begin{figure}[htp]
		\includegraphics[width=1\columnwidth]{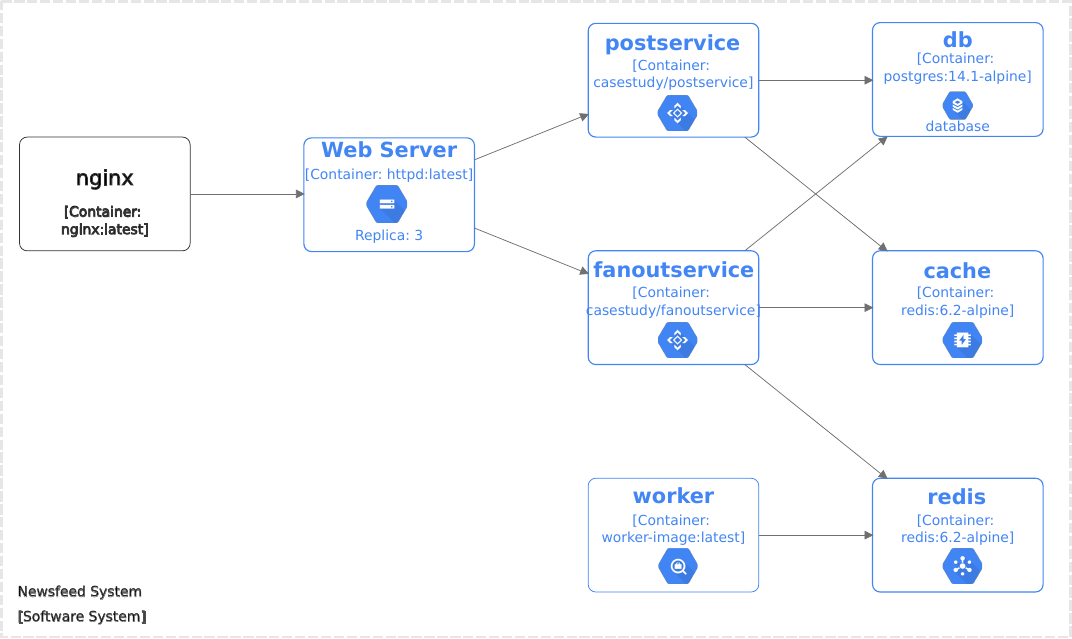}
		\caption{High-level architecture of a newsfeed system. }
		\label{fig:newsfeed-system-diagram}
	\end{figure}

Finally, from the DaC model, the Structurzr tool generates the diagram in Figure \ref{fig:newsfeed-system-diagram}.

\begin{lstlisting}[caption={Newsfeed System Diagram-as-code model}, label={lst:newsfeed-dac-model}]
	workspace {
		model {
			softwareSystem = softwareSystem "Newsfeed System" {
				nginx = container "nginx" "" "nginx:latest"{}
				webserver = container "Web Server"  "Replica: 3" "httpd:latest" "Replicas: 3"{}
				db = container "db" "database" "postgres:14.1-alpine"  {}
				cache = container "cache" "" "redis:6.2-alpine"{}
				
				nginx -> webserver1 "depends_on"
			}
		}
	\end{lstlisting}

\section{Discussion}
\label{sec:discussion}

We compared the architectural diagram of Xu \cite{xu2020} (Figure \ref{fig:newsfeed-system-adhoc-xu-diagram}) with the architectural diagrams rendered in our case study (Figure \ref{fig:newsfeed-system-diagram}), and we discuss our observation as follows.

First, we note that the architecture diagram contains all the elements of the system descriptor. Since our transformation function (consisting of P1, P2, and rendering processes) has succeeded in generating a diagram $\delta$ from a system descriptor $\sigma$, such that $\delta = f(\sigma)$, then we can say that the rendered architectural diagram is consistent.

\begin{tcolorbox}[colframe=gray!50, coltitle=black]
	\textbf{Observation 1: }The rendered architectural diagram accurately describes the system descriptor elements and their relationships.
\end{tcolorbox}

Analyzing the diagram in terms of the guidelines presented in Section \ref{sec:survey-results}, we can see that all guidelines have been met, except for guideline number 4, since this type of information is missing from the system description. In fact, some elements of the original architecture diagram in Figure \ref{fig:newsfeed-system-adhoc-xu-diagram} do not appear in the rendered diagram, such as user elements (web browser and mobile app) or DNS elements. This is because these elements are external nodes that interact with the system and, for this reason, are not included in the system descriptors, regardless of the technology used.

However, we believe that external elements play an essential role in understanding systems because they relate the system to external entities, provide an overview of how the system interacts with external elements, and contribute to greater consistency in the architectural diagram.

\begin{tcolorbox}[colframe=gray!50, coltitle=black]
	\textbf{Observation 2: }External nodes play an important role in the diagram comprehension and should be rendered.
\end{tcolorbox}

Finally, we found it challenging to map the system descriptor to the meta-descriptor because it was difficult to identify the essential elements of the meta-descriptor (components, component types, and their relationships) in the system descriptor. For example, Kubernetes does not inherently support the use of a term like \textit{depends\_on} in its descriptor, which makes it difficult to map relationships between components. Similarly, in Docker Compose and Kubernetes, it was not possible to identify the existence of a specific term that informs about the type of component. 

A solution to this problem in Kubernetes is for the developer to create custom terms for this purpose. In Docker Compose, on the other hand, a solution could be to use another term, such as container\_name, to identify the type of component. Nevertheless, these solutions depend on the quality and details of the system descriptor document written by the developer and would make it difficult to automate the transformation process.

\begin{tcolorbox}[colframe=gray!50, coltitle=black]
	\textbf{Observation 3: }The difficulties encountered in mapping elements from system descriptors to meta-descriptors make the P1 process more challenging and complex to automate.
\end{tcolorbox}

\subsection{The Consistency of Architectural Diagrams}
This section gives answers to research question: \textit{Does building software architecture diagrams using system descriptors improve the consistency of architecture diagrams?}

In light of the preceding observations, we have found that system descriptors generally lead to consistent diagrams only to a limited extent. However, based on the evidence shown in this paper, we observe the potential of system descriptors to create architectural diagrams.

We suggest adding annotations in that system descriptors to avoid the difficulties encountered in mapping descriptors. These annotations would highlight missing or difficult-to-assign information from the descriptors (e.g., component type and dependency relationship). Developers should insert the annotations as comments in the code to avoid disturbing the processing of the system descriptors. Finally, the P1 process of our approach would read these annotations and use the information to complete the mapping of the components and their relationships.

We justify these adaptations' costs by the benefits of this approach to creating architectural diagrams. The advantages of creating architectural diagrams from descriptors are: 
\begin{enumerate}
	\item the architecture diagrams can represent all descriptor elements.
	\item the architecture diagrams remain synchronized with the evolution of the descriptors and consequently of the system.
	\item the architecture diagrams avoid inconsistency errors due to system evolution, as diagrams synchronize with system descriptors.
\end{enumerate}

\subsection{Threats to Validity}

This section identifies and discusses threats that could affect the results' validity. We identify the following threats.

Regarding the external validity, some aspects could affect the general applicability of our results. Our case study refers to modeling a single system using only one system descriptor provider applying the approach. For this reason, we cannot generalize the results. Nevertheless, we believe the results are valid as they provide relevant insights into the potential of descriptors to build architectural diagrams. 

For example, the architectural diagram rendered in the case study could clearly represent all system descriptor elements. In addition, Wurstser showed that it is possible to map several other system descriptor providers to the EDMM meta-model, such as Puppet, Chef, Ansible, Terraform, OpenStack, SaltStack, AWS, Azure, Kubernetes, and Docker Compose \cite{wurster2019}.

Since we are using the EDMM meta-model to develop the taxonomy of system descriptor terms, we believe it is possible to generate a meta-descriptor from each of the above descriptors. We intend to explore these issues in more detail in future work.

\section{Conclusion}
\label{sec:conclusions}

In this paper, we present our approach to applying system descriptors to improve the consistency of architectural diagrams. First, we introduced relevant concepts related to this work and presented related work. Then, we surveyed the problems with architecture diagrams. From the analysis of the survey responses, we have compiled a set of guidelines for creating architecture diagrams. Next, we presented our approach, introducing a taxonomy for system descriptor concepts and a function for converting system descriptors into architectural diagrams. Further, we presented a case study evaluating our approach.

Our case study shows that the rendered architecture diagram accurately describes the system descriptor elements and their relationships. However, we found that some external system elements present in the original architecture diagram, such as user elements, do not appear in the rendered diagram because the descriptor does not include these elements. We believe that external nodes play an essential role in understanding the diagram and, therefore, should be rendered. In addition, we found that the difficulty in mapping elements from system descriptors to meta-descriptors makes the P1 process more difficult and complex to automate. Additionally, some details in the descriptors, such as service ports, could be incorporated into architectural diagrams as an additional information layer. Also, other details, such as passwords, were not mapped in the generated architectural diagram. Such information describes details that are out of the scope of architectural diagrams. Unmapped elements of a system descriptor can be specified as a subset of the domain of that mapping.

Based on these insights, we have found that system descriptors generally lead to consistent diagrams only to a limited extent. However, the case study's observations indicate that the proposed approach is promising and demonstrates that system descriptors have the potential to create more consistent architectural diagrams.

Our case study only uses a single system descriptor provider: Docker Compose. In future work, we intend to carry out studies with other descriptor providers. Also, we intend to implement a tool to evaluate our hypotheses more deeply in a controlled experiment.

\bibliographystyle{abbrv}
\bibliography{main}

\end{document}